\documentclass[twocolumn,showpacs,preprintnumbers]{revtex4}%
\usepackage{amssymb}
\usepackage{amsmath}
\usepackage{graphicx}
\usepackage{dcolumn}
\usepackage{bm}
\usepackage{amsfonts}%
\begin{document}
\title{Mass Detection with Nonlinear Nanomechanical Resonator}
\author{Eyal Buks}
\affiliation{Department of Electrical Engineering, Technion, Haifa 32000 Israel}
\author{Bernard Yurke}
\affiliation{Bell Laboratories, Lucent Technologies, 600 Mountain Avenue, Murray Hill, NJ 07974}
\date{\today }

\begin{abstract}
Nanomechanical resonators having small mass, high resonance frequency and low
damping rate are widely employed as mass detectors. We study the performances
of such a detector when the resonator is driven into a region of nonlinear
oscillations. We predict theoretically that in this region the system acts as
a phase-sensitive mechanical amplifier. This behavior can be exploited to
achieve noise squeezing in the output signal when homodyne detection is
employed for readout. We show that mass sensitivity of the device in this
region may exceed the upper bound imposed by thermomechanical noise upon the
sensitivity when operating in the linear region. On the other hand, we show
that the high mass sensitivity is accompanied by a slowing down of the
response of the system to a change in the mass.

\end{abstract}
\pacs{42.50.Dv, 05.45.-a}
\maketitle





\section{Introduction}

Nano-electro-mechanical systems (NEMS) serve in a variety of applications as
sensors and actuators. Recent studies have demonstrated ultra-sensitive mass
sensors based on NEMS
\cite{Wachter_3662,Lang_383,Hu_427,Ilic_450,Fritz_316,Ilic_2825,Subramanian_385,Ilic_2604,Ilic_3694,Ilic_925,Ekinci_4469,Ghatkesar_1060,Ghatnekar_98}%
. Such sensors promise a broad range of applications, from ultra-sensitive
mass spectrometers that can be used to detect hazardous molecules, through
biological applications at the level of a single DNA base-pair, to the study
of fundamental questions such as the interaction of a single pair of
molecules. In these devices mass detection is achieved by monitoring the
resonance frequency $\omega_{0}$ of one of the modes of a nanomechanical
resonator. The dependence of $\omega_{0}$ on the effective mass $m$ allows for
sensitive detection of additional mass being adsorbed on the surfaces of the
resonator. In such mass detectors the adsorbent molecules are anchored to the
resonator surface either by Van der-Waals interaction, or by covalent bonds to
linker molecules that are attached to the surface. Various analytes were used
in those experiments, including alcohol and explosive gases, biomolecules,
single cells, DNA molecules, and alkane chains. Currently, the smallest
detectable mass change is $\delta m\simeq0.4\times10^{-21}%
\operatorname{kg}%
$ \cite{Ilic_3694}, achieved by using a $4%
\operatorname{\mu m}%
$ long silicon beam with a resonance frequency $\omega_{0}/2\pi=10%
\operatorname{MHz}%
$, a quality factor $Q$ of about $2,500$, and total mass $m\simeq
5\times10^{-16}%
\operatorname{kg}%
$. In a recent experiment Ilic \textit{et al}. \cite{Ilic_925} succeeded to
measure a single DNA molecule of about $1,600$ base pairs, which corresponds
to $\delta m\simeq1.6\times10^{-21}%
\operatorname{kg}%
$, by using a silicon nitride cantilever, and employing an optical detection scheme.

In general, any detection scheme employed for monitoring the mass can be
characterized by two important figures of merit. The first is the minimum
detectable change in mass $\delta m$. This parameter is determined by the
responsivity, which is defined as the derivative of the average output signal
$\left\langle X\left(  t\right)  \right\rangle $ of the detector with respect
to the mass $m$, the noise level, which is usually characterized by the
spectral density of $X\left(  t\right)  $, and by the averaging time $\tau$
employed for measuring the output signal $X\left(  t\right)  $. The second
figure of merit is the ring-down time $t_{\mathrm{RD}}$, which is a measure of
the time width of the step in $X\left(  t\right)  $ due to a sudden change in
$m$.

A number of factors affect the minimum detectable mass $\delta m$ and the
ring-down time $t_{\mathrm{RD}}$ of mass detectors, based on nanomechanical
resonators. Recent studies \cite{Ekinci_2682,Cleland_235} have shown that if
measurement noise is dominated by thermomechanical fluctuations the following hold%

\begin{equation}
\frac{\delta m}{m}=2\left(  \frac{2\pi}{Q\omega_{0}\tau}\frac{k_{B}T}{U_{0}%
}\right)  ^{1/2}\ , \label{delta m linear}%
\end{equation}

where $k_{B}T$ is the thermal energy, $U_{0}$ is the energy stored in the
resonator, and $\tau$ is the measurement averaging time, and the ring-down
time is given by%

\begin{equation}
t_{\mathrm{RD}}=\frac{Q}{\omega_{0}}\ . \label{t_RD linear}%
\end{equation}

Eq. (\ref{delta m linear}) indicates that nanomechanical resonators having
small $m$ and high $\omega_{0}$ may allow high mass sensitivity (small $\delta
m$). Further enhancement in the sensitivity can be achieved by increasing $Q$
however, this will be accompanied by an undesirable increase in the ring-down
time, namely, slowing down the response of the system to changes in $m$.
Moreover, Eq. (\ref{delta m linear}) apparently suggests that unlimited
reduction in $\delta m$ can be achieved by increasing $U_{0}$ by means of
increasing the drive amplitude. Note however that Eq. (\ref{delta m linear}),
which was derived by assuming the case of linear response, is not applicable
in the nonlinear region. Thus, in order to characterize the performances of
the system when nonlinear oscillations are excited by an intense drive, one
has to generalize the analysis by taking nonlinearity into account. From a
more general point of view, such a generalization is interesting because it
provides some insight onto the question of what is the range of applicability
of the fluctuation-dissipation theorem for nonlinear systems
\cite{Wang_025008}.

In the present paper we generalize Eqs. (\ref{delta m linear}) and
(\ref{t_RD linear}) and extend their range of applicability by taking into
account nonlinearity in the response of the resonator to lowest order.
Practically, characterizing the performances of nanomechanical mass detectors
in the nonlinear region is important since in many cases, when a displacement
detector with a sufficiently high sensitivity is not available, the
oscillations of the system in the linear regime cannot be monitored, and
consequently operation is possible only in the region of nonlinear
oscillations. Another possibility for exploiting nonlinearity for enhancing
mass sensitivity was recently studied theoretically by Cleland
\cite{Cleland_235}, who has considered the case where the mechanical resonator
is excited parametrically.

When nonlinearity is taken into account to lowest order the resonator's
dynamics can be described by the Duffing equation of motion \cite{Mechanics}.
A Duffing resonator may exhibit bistability when driven by an external
periodic force with amplitude $p$ exceeding some critical value $p_{c}$.
Figure \ref{Duffing} below shows the calculated response vs. drive frequency
$\omega_{p}$ of a Duffing resonator excited by a driving force with (b)
sub-critical $p=p_{c}/2$ (c) critical $p=p_{c}$, and (d) over-critical
$p=2p_{c}$ amplitude. The range of bistability in the $\left(  \omega
_{p},p\right)  $ plane is seen in Fig. \ref{Duffing} (a). As was shown in Ref.
\cite{Squeezing_Yurke05}, high responsivity can be achieved when driving the
resonator close to the edge of the bistability region
\cite{Wiesenfeld_629,Dykman_1198,Krommer_101,Savel'ev_056136,Chan_0603037},
where the slope of the response vs. frequency curve approaches infinity. Note
however that in the same region of operation an undesirable slowing down
occurs, namely $t_{\mathrm{RD}}$ can become much longer than its value in the
linear region, which is given by Eq. (\ref{t_RD linear}).%

\begin{figure}
[ptb]
\begin{center}
\includegraphics[
trim=0.233662in 0.471040in 2.361261in 2.125181in,
height=4.6527in,
width=3.2742in
]%
{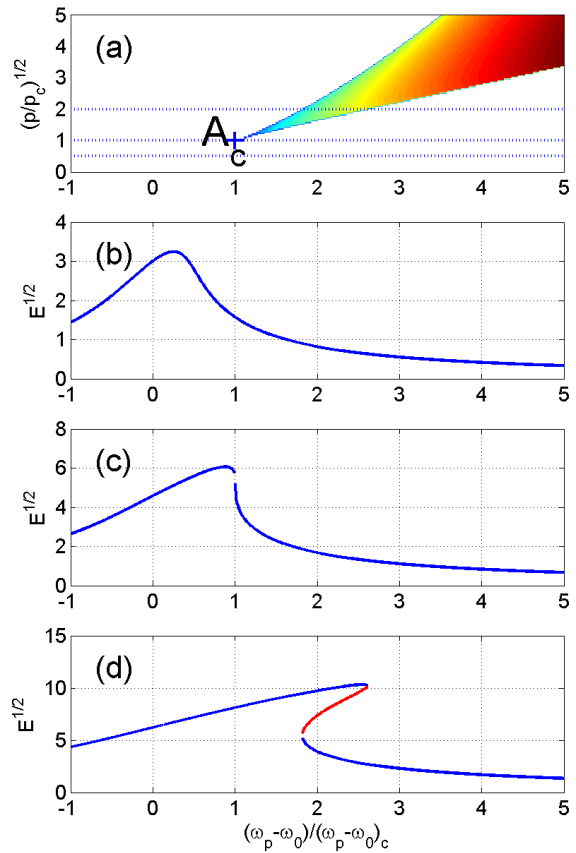}%
\caption{(Color online) Response of a driven Duffing resonator. Panel (a)
shows the bistable region in the $\left(  \omega_{p},p\right)  $ plane. The
response vs. frequency is shown in panel (b), (c) and (d) for sub-critical,
critical, and over-critical driving force respectively.}%
\label{Duffing}%
\end{center}
\end{figure}

The detector's performances depend in general on the detection scheme which is
being employed. Here we consider the case of a homodyne detection scheme
\cite{Squeezing_Yurke05}, where the output signal of a displacement detector
monitoring the mechanical motion of the resonator is mixed with a local
oscillator at the frequency of the driving force and with an adjustable phase
$\phi_{\mathrm{LO}}$. In the nonlinear regime of operation the device acts as
a phase sensitive intermodulation amplifier \cite{Almog_213509}. Consequently,
noise squeezing occurs in this regime, as was recently demonstrated
experimentally in Ref. \cite{Almog_0607055}, namely, the spectral density of
the output signal at the IF port of the mixer depends on $\phi_{\mathrm{LO}}$
periodically \cite{Rugar_699}.

To optimize the operation of the system in the nonlinear region it is
important to understand the role played by damping. In this region, in
addition to linear damping, also nonlinear damping may affect the device
performances. Our theoretical analysis \cite{Squeezing_Yurke05} shows that
instability in a Duffing resonator is accessible only when the nonlinear
damping is sufficiently small. Moreover, a fit between theory and experimental
results allows extracting the nonlinear damping rate. By employing such a fit
it was found in Ref. \cite{Zaitsev_0503130} that nonlinear damping can play a
significant role in the dynamics in the nonlinear region, and thus we take it
into account in our analysis.

The paper is organized as follows. In section II the Hamiltonian of the driven
Duffing resonator is introduced. The equations of motion of the system are
derived in section III and linearized in section IV. The basins of attraction
of the system are presented in section V. The ring-down time is estimated in
section VI, whereas the case of homodyne detection is discussed in section
VII. The calculation of the spectral density of the output signal of the
homodyne detector, which is presented in section VIII, allows to calculate the
minimum detectable mass in section IX. We conclude by comparing our findings
with the linear case in section X.

\section{Hamiltonian}

Consider a nonlinear mechanical resonator of mass $m$, resonance frequency
$\omega_{0}$, damping rate $\gamma$, nonlinear Kerr constant $K$, and
nonlinear damping rate $\gamma_{3}$. The resonator is driven by harmonic force
at frequency $\omega_{p}$. The complex amplitude of the force $f$ is written
as%
\begin{equation}
f=-2im\omega_{p}x_{0}p^{1/2}e^{i\phi_{p}}\ , \label{f=}%
\end{equation}

where $p$ is positive real, $\phi_{p}$ is real, and $x_{0}$ is give by%

\begin{equation}
x_{0}=\sqrt{\frac{\hbar}{2m\omega_{0}}}\ .
\end{equation}

The Hamiltonian of the system is given by \cite{Squeezing_Yurke05}%
\begin{equation}
H=H_{1}+H_{a_{2}}+H_{a_{3}}+H_{c2}+H_{c3}\ ,
\end{equation}

where $H_{1}$ is the Hamiltonian for the driven nonlinear resonator
\begin{align}
H_{1}  &  =\hbar\omega_{0}A^{\dagger}A+\frac{\hbar}{2}KA^{\dagger}A^{\dagger
}AA\nonumber\\
&  +\hbar p^{1/2}\left(  ie^{i\left(  \phi_{p}-\omega_{p}t\right)  }%
A^{\dagger}-ie^{-i\left(  \phi_{p}-\omega_{p}t\right)  }A\right)
\ .\nonumber\\
&
\end{align}

The resonator's creation and annihilation operators satisfy the following
commutation relation%

\begin{equation}
\left[  A,A^{\dagger}\right]  =AA^{\dagger}-A^{\dagger}A=1\ .
\end{equation}

The Hamiltonians $H_{a2}$ and $H_{a3}$ associated with both baths are given
by
\begin{equation}
H_{a2}=\int\mathrm{d}\omega\hbar\omega a_{2}^{\dagger}\left(  \omega\right)
a_{2}\left(  \omega\right)  \ ,
\end{equation}

\begin{equation}
H_{a3}=\int\mathrm{d}\omega\hbar\omega a_{3}^{\dagger}\left(  \omega\right)
a_{3}\left(  \omega\right)  \ .
\end{equation}

The Hamiltonian $H_{c2}$ linearly couples the bath modes $a_{2}\left(
\omega\right)  $ to the resonator mode $A$
\begin{equation}
H_{c2}=\hbar\sqrt{\frac{\gamma}{\pi}}\int\mathrm{d}\omega\left[  e^{i\phi_{2}%
}A^{\dagger}a_{2}\left(  \omega\right)  +e^{-i\phi_{2}}a_{2}^{\dagger}\left(
\omega\right)  A\right]  \ ,
\end{equation}

whereas $H_{c3}$ describes two-phonon absorptive coupling of the resonator
mode to the bath modes $a_{3}\left(  \omega\right)  $ in which two resonator
phonons are destroyed for every bath phonon created
\begin{equation}
H_{c3}=\hbar\sqrt{\frac{\gamma_{3}}{\pi}}\int\mathrm{d}\omega\left[
e^{i\phi_{3}}A^{\dagger}A^{\dagger}a_{3}(\omega)+e^{-i\phi_{3}}a_{3}^{\dagger
}(\omega)AA\right]  \ .
\end{equation}

Both phase factors $\phi_{2}$ and $\phi_{3}$ are real. The bath modes are
boson modes, satisfying the usual Bose commutation relations%

\begin{align}
\lbrack a_{n}(\omega),a_{n}^{\dagger}(\omega^{\prime})]  &  =\delta\left(
\omega-\omega^{\prime}\right)  \ ,\\
\ [a_{n}(\omega),a_{n}(\omega^{\prime})]  &  =0\ .
\end{align}

\section{Equations of Motion}

We now generate the Heisenberg equations of motion according to
\begin{equation}
i\hbar\frac{dO}{dt}=\left[  O,H\right]  \ ,
\end{equation}
where $O$ is an operator and $H$ is the total Hamiltonian%

\begin{align}
i\frac{\mathrm{d}A}{\mathrm{d}t}  &  =\omega_{0}A+KA^{\dagger}AA+ip^{1/2}%
e^{i\phi_{p}}e^{-i\omega_{p}t}\nonumber\\
&  +\sqrt{\frac{\gamma}{\pi}}e^{i\phi_{2}}\int\mathrm{d}\omega a_{2}\left(
\omega\right)  +2\sqrt{\frac{\gamma_{3}}{\pi}}e^{i\phi_{3}}A^{\dagger}%
\int\mathrm{d}\omega a_{3}(\omega)\ ,\nonumber\\
&
\end{align}

\begin{equation}
\frac{\mathrm{d}a_{2}\left(  \omega\right)  }{\mathrm{d}t}=-i\omega
a_{2}\left(  \omega\right)  -i\sqrt{\frac{\gamma}{\pi}}e^{-i\phi_{2}}A\ ,
\end{equation}

\begin{equation}
\frac{\mathrm{d}a_{3}\left(  \omega\right)  }{\mathrm{d}t}=-i\omega
a_{3}\left(  \omega\right)  -i\sqrt{\frac{\gamma_{3}}{2\pi}}e^{-i\phi_{3}%
}AA\ .
\end{equation}

Using the standard method of Gardiner and Collett \cite{Gardiner_3761}, and
employing a transformation to a reference frame rotating at angular frequency
$\omega_{p}$%

\begin{equation}
A=Ce^{-i\omega_{p}t}\ ,
\end{equation}

yield the following equation for the operator $C$%

\begin{equation}
\frac{\mathrm{d}C}{\mathrm{d}t}+\Theta=F\left(  t\right)  \ , \label{dC/dt}%
\end{equation}

where%

\begin{equation}
\Theta=\left[  \gamma+i\left(  \omega_{0}-\omega_{p}\right)  +\left(
iK+\gamma_{3}\right)  C^{\dagger}C\right]  C-p^{1/2}e^{i\phi_{p}}\ .
\label{Theta(C,C+)}%
\end{equation}

The noise term $F\left(  t\right)  $ is given by%

\begin{equation}
F=-i\sqrt{2\gamma}e^{i\phi_{2}}a_{2}^{in}e^{i\omega_{p}t}-i2\sqrt{\gamma_{3}%
}e^{i\phi_{3}}C^{\dagger}a_{3}^{in}e^{2i\omega_{p}t}\ ,
\end{equation}

where%

\begin{equation}
a_{2}^{in}\left(  t\right)  =\frac{1}{\sqrt{2\pi}}\int\mathrm{d}\omega
e^{-i\omega\left(  t-t_{0}\right)  }a_{2}\left(  t_{0},\omega\right)  \ ,
\end{equation}%
\begin{equation}
a_{3}^{in}\left(  t\right)  =\frac{1}{\sqrt{2\pi}}\int\mathrm{d}\omega
e^{-i\omega\left(  t-t_{0}\right)  }a_{3}\left(  t_{0},\omega\right)  \ .
\end{equation}

Note that in the noiseless case, namely when $F=0$, the equation of motion for
the displacement $x$ of the vibrating mode can be written as%

\begin{align}
&  \frac{\mathrm{d}^{2}x}{\mathrm{d}t^{2}}+2\gamma\left[  1+\frac{\gamma_{3}%
}{3\gamma}\left(  \frac{x}{x_{0}}\right)  ^{2}\right]  \frac{\mathrm{d}%
x}{\mathrm{d}t}+\omega_{0}^{2}\left[  1+\frac{2K}{3\omega_{0}}\left(  \frac
{x}{x_{0}}\right)  ^{2}\right]  x\nonumber\\
&  =\frac{f}{m}e^{-i\omega_{p}t}+c.c.\ .\nonumber\\
&
\end{align}

\section{Linearization}

Let $C=C_{m}+c$, where $C_{m}$ is a complex number for which%
\begin{equation}
\Theta\left(  C_{m},C_{m}^{\ast}\right)  =0\ , \label{Theta(C_m,C_m^*)}%
\end{equation}
namely, $C_{m}$ is a steady state solution of Eq. (\ref{dC/dt}) for the
noiseless case $F=0$. When the noise term $F$ can be considered as small, one
can find an equation of motion for the fluctuation around $C_{m}$ by
linearizing Eq. (\ref{dC/dt})%

\begin{equation}
\frac{\mathrm{d}c}{\mathrm{d}t}+Wc+Vc^{\dagger}=F\ , \label{dc/dt}%
\end{equation}

where%

\begin{equation}
W=\left.  \frac{\partial\Theta}{\partial C}\right\vert _{C=C_{m}}%
=\gamma+i\left(  \omega_{0}-\omega_{p}\right)  +2\left(  iK+\gamma_{3}\right)
C_{m}^{\ast}C_{m}\ , \label{W}%
\end{equation}

and%

\begin{equation}
V=\left.  \frac{\partial\Theta}{\partial C^{\dagger}}\right\vert _{C=C_{m}%
}=\left(  iK+\gamma_{3}\right)  C_{m}^{2}\ . \label{V}%
\end{equation}

\subsection{Mean-Field Solution}

Using the notation%

\begin{equation}
C_{m}=E^{1/2}e^{i\phi_{m}}\ , \label{C_m=E^(1/2)*exp(i*phi_m)}%
\end{equation}

where $E$ is positive and $\phi_{m}\ $is real, Eq. (\ref{Theta(C_m,C_m^*)}) reads%

\begin{equation}
\left[  \gamma+i\left(  \omega_{0}-\omega_{p}\right)  +\left(  iK+\gamma
_{3}\right)  E\right]  E^{1/2}e^{i\phi_{m}}=p^{1/2}e^{i\phi_{p}}\ .
\label{eq for C_m}%
\end{equation}

Multiplying each side by its complex conjugate yields%

\begin{equation}
\left[  \left(  \gamma+\gamma_{3}E\right)  ^{2}+\left(  \omega_{0}-\omega
_{p}+KE\right)  ^{2}\right]  E=p\ . \label{eq for E}%
\end{equation}

Finding $E$ by solving the cubic polynomial Eq. (\ref{eq for E}) allows
calculating $C_{m}$ using Eq. (\ref{eq for C_m}).

Taking the derivative of Eq. (\ref{eq for E}) with respect to the drive
frequency $\omega_{p}$, one finds
\begin{equation}
\frac{\partial E}{\partial\omega_{p}}=\frac{2(\omega_{0}-\omega_{p}%
+KE)E}{\left\vert W\right\vert ^{2}\left(  1-\zeta^{2}\right)  }\ ,
\end{equation}

where%

\begin{equation}
\zeta=\left\vert \frac{V}{W}\right\vert \ .
\end{equation}

Similarly for the drive amplitude $p$%

\begin{equation}
\frac{\partial E}{\partial p}=\frac{1}{\left\vert W\right\vert ^{2}\left(
1-\zeta^{2}\right)  }\ .
\end{equation}

Note that, as will be shown below, the value $\zeta=1$ occurs along the edge
of the bistability region.

\subsection{The bifurcation point}

At the bifurcation point, namely at the onset of bistability, the following
holds%
\begin{equation}
\frac{\partial\omega_{p}}{\partial E}=\frac{\partial^{2}\omega_{p}}{\partial
E^{2}}=0\ .
\end{equation}
Such a point occurs only if the nonlinear damping is sufficiently small
\cite{Squeezing_Yurke05}, namely, only when the following condition holds%
\begin{equation}
|K|>\sqrt{3}\gamma_{3}\ .
\end{equation}
At the bifurcation point the drive frequency and amplitude are given by%
\begin{equation}
\left(  \omega_{p}-\omega_{0}\right)  _{c}=\gamma\frac{K}{|K|}\left[
\frac{4\gamma_{3}|K|+\sqrt{3}\left(  K^{2}+\gamma_{3}^{2}\right)  }%
{K^{2}-3\gamma_{3}^{2}}\right]  \ , \label{omega_c}%
\end{equation}%
\begin{equation}
p_{c}=\frac{8}{3\sqrt{3}}\frac{\gamma^{3}(K^{2}+\gamma_{3}^{2})}{\left(
|K|-\sqrt{3}\gamma_{3}\right)  ^{3}}\ , \label{b^in_1c}%
\end{equation}

and the resonator mode amplitude is%

\begin{equation}
E_{c}=\frac{2\gamma}{\sqrt{3}\left(  |K|-\sqrt{3}\gamma_{3}\right)  }\ .
\label{B^2_c}%
\end{equation}

\section{Basins of Attraction}

In the bistable region Eq. (\ref{Theta(C_m,C_m^*)}) has 3 different solutions,
labeled as $C_{1},$ $C_{2}$ and $C_{3}$, where both stable solutions $C_{1}$
and $C_{3}$ are attractors, and the unstable solution $C_{2}$ is a saddle
point. The bistable region $\Lambda$\ in the plane of parameters $\left(
\omega_{p},p\right)  $ is seen in the colormap in Fig. \ref{Duffing} (a). The
Kerr constant in this example is $K/\omega_{0}=0.001$, and the damping
constants are $\gamma/\omega_{0}=0.02$, $\gamma_{3}=0.1K/\sqrt{3}$. The color
in the bistable region $\Lambda$ indicates the difference $\left\vert
C_{3}\right\vert ^{2}-\left\vert C_{1}\right\vert ^{2}$. The bifurcation point
at $\omega_{p}-\omega_{0}=\left(  \omega_{p}-\omega_{0}\right)  _{c}$ and
$p=p_{c}$ is labeled as $A_{c}$ in the figure.

\begin{figure}
[ptb]
\begin{center}
\includegraphics[
height=4.7106in,
width=3.0355in
]%
{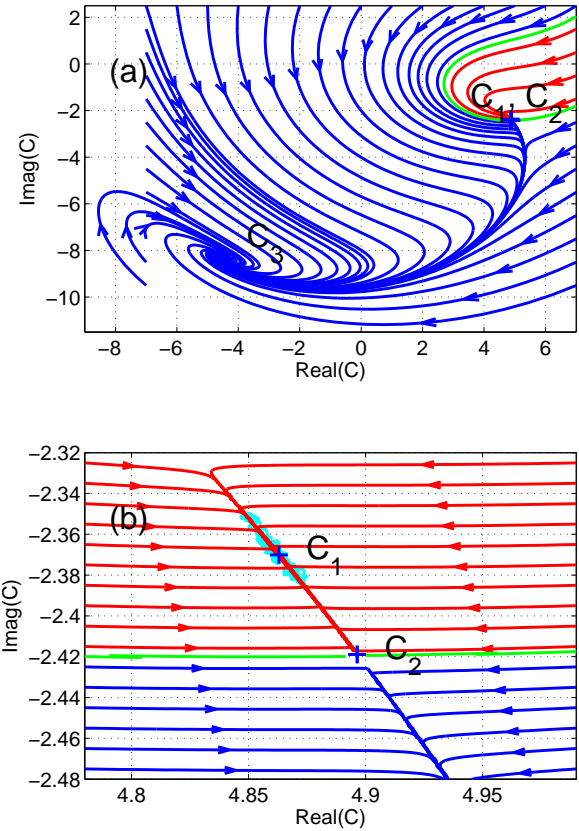}%
\caption{(Color online) Flow lines obtained by integrating Eq. (\ref{dC/dt})
for the noiseless case $F=0$. The points $C_{1}$ and $C_{3}$ are attractors,
and $C_{2}$ is a saddle point. The green line is the seperatrix, namely the
boundary between the basins of attraction of both attractors. Panel (a) shows
a wide view, whereas panel (b) shows a closer view of the region near $C_{1}$
and $C_{2}$. The cyan line near the attractor $C_{1}$ in panel (b)
demonstrates random motion in the presence of noise.}%
\label{Flow}%
\end{center}
\end{figure}

Figure \ref{Flow} (a) shows some flow lines obtained by integrating Eq.
(\ref{dC/dt}) numerically for the noiseless case $F=0$. The red and blue lines
represent flow toward the attractors at $C_{1}$ and $C_{3}$ respectively. The
green line is the seperatrix, namely the boundary between the basins of
attraction of the attractors at $C_{1}$ and $C_{3}$. A closer view of the
region near $C_{1}$ and $C_{2}$ is given in Fig. \ref{Flow} (b). This figure
shows also, an example of a random motion near the attractor at $C_{1}$ (seen
as a cyan line). The line was obtained by numerically integrating Eq.
(\ref{dC/dt}) with a non vanishing fluctuating force $F$. The random walk
demonstrates noise squeezing (to be further discussed below), where the
fluctuations obtain their largest and smallest values along the directions of
the local principle axes (see appendix).%

\section{Ring-Down Time}

The solution of the equation of motion (\ref{dc/dt}) was found in Ref.
\cite{Squeezing_Yurke05}%

\begin{equation}
c\left(  t\right)  =\int_{-\infty}^{\infty}\mathrm{d}t^{\prime}G\left(
t-t^{\prime}\right)  \Gamma\left(  t^{\prime}\right)  \ ,
\end{equation}
where%

\begin{equation}
\Gamma\left(  t\right)  =\frac{\mathrm{d}F\left(  t\right)  }{dt}+W^{\ast
}F\left(  t\right)  -VF^{\dagger}\left(  t\right)  \ .
\end{equation}

The propagator is given by%

\begin{equation}
G\left(  t\right)  =u\left(  t\right)  \frac{e^{-\lambda_{0}t}-e^{\lambda
_{1}t}}{\lambda_{1}-\lambda_{0}}\ ,
\end{equation}
where $u(t)$ is the unit step function
\begin{equation}
u(t)=\left\{
\begin{array}
[c]{ll}%
1, & t>0\\
1/2, & t=0\\
0, & t<0
\end{array}
\right.  \ ,
\end{equation}

and $\lambda_{0}$ and $\lambda_{1}$ are the eigenvalues of the homogeneous
equation, which satisfy%

\begin{equation}
\lambda_{0}+\lambda_{1}=2W^{\prime}\ , \label{lam_0+lam_1}%
\end{equation}

\begin{equation}
\lambda_{0}\lambda_{1}=|W|^{2}-|V|^{2}\ , \label{lam_0*lam_1}%
\end{equation}

where $W^{\prime}$ is the real part of $W$. Thus one has%

\begin{equation}
\lambda_{0,1}=W^{\prime}\left(  1\pm\sqrt{1+\frac{|W|^{2}}{\left(  W^{\prime
}\right)  ^{2}}\left(  \zeta^{2}-1\right)  }\right)  \ ,
\end{equation}

or%

\begin{equation}
\lambda_{0,1}=\gamma+2\gamma_{3}E\pm\sqrt{\left(  K^{2}+\gamma_{3}^{2}\right)
E^{2}-\left(  \omega_{0}-\omega_{p}+2KE\right)  ^{2}}\ .
\end{equation}

We chose to characterize the ring-down time scale as%

\begin{equation}
t_{\mathrm{RD}}=\left(  \lambda_{0}\lambda_{1}\right)  ^{-1/2}=\frac
{1}{|W|\sqrt{1-\zeta^{2}}}\ . \label{t_RD nonlinear}%
\end{equation}

Note that in the limit $\zeta\rightarrow1$ slowing down occurs and
$t_{\mathrm{RD}}\rightarrow\infty$. This limit corresponds to the case of
operating the resonator near a jump point close to the edge of the bistability region.

\section{Homodyne Detection}

Consider the case where homodyne detection is employed for readout. In this
case the output signal of a displacement detector monitoring the mechanical
motion is mixed with a local oscillator at the same frequency as the frequency
of the pump $\omega_{p}$ and having an adjustable phase $\phi_{\mathrm{LO}}$
($\phi_{\mathrm{LO}}$ is real). The local oscillator is assumed to be
noiseless. The output signal of the homodyne detector is proportional to%

\begin{equation}
X_{\phi_{\mathrm{LO}}}\left(  t\right)  =e^{i\phi_{\mathrm{LO}}}C\left(
t\right)  +e^{-i\phi_{\mathrm{LO}}}C^{\dag}\left(  t\right)  \ .
\end{equation}

For the stationary case of a fixed mass $m$ the time varying signal
$X_{\phi_{\mathrm{LO}}}\left(  t\right)  $ can be characterized by its average%

\begin{equation}
X_{0}=\left\langle X_{\phi_{\mathrm{LO}}}\left(  t\right)  \right\rangle \ ,
\end{equation}

and by its time auto-correlation function%

\begin{equation}
K\left(  t^{\prime}-t\right)  =\left\langle \left[  X_{\phi_{\mathrm{LO}}%
}\left(  t\right)  -X_{0}\right]  \left[  X_{\phi_{\mathrm{LO}}}\left(
t^{\prime}\right)  -X_{0}\right]  \right\rangle \ .
\end{equation}

The correlation function is expected to be an even function of $t^{\prime}-t$
with a maximum at $t^{\prime}-t=0$. The correlation time characterizes the
width of that peak. Consider a measurement in which $X_{\phi_{\mathrm{LO}}%
}\left(  t\right)  $ is continuously monitored in the time interval $\left[
0,\tau\right]  $. Let $X_{\tau}$ be an estimator of the average value of
$X_{\phi_{\mathrm{LO}}}\left(  t\right)  $%

\begin{equation}
X_{\tau}=\frac{1}{\tau}\int_{0}^{\tau}\mathrm{d}t\ X_{\phi_{\mathrm{LO}}%
}\left(  t\right)  \ .
\end{equation}

Clearly $X_{\tau}$ is unbiased, and its variance is given by%

\begin{equation}
\left\langle \left(  X_{\tau}-X_{0}\right)  ^{2}\right\rangle =\frac{1}%
{\tau^{2}}\int_{0}^{\tau}\mathrm{d}t\int_{0}^{\tau}\mathrm{d}t^{\prime
}\ K\left(  t^{\prime}-t\right)  \ .
\end{equation}

Assuming the case where the measurement time $\tau$ is much longer than the
correlation time. For this case one can employ the approximation%

\begin{equation}
\left\langle \left(  X_{\tau}-X_{0}\right)  ^{2}\right\rangle =\frac{1}{\tau
}\int_{-\infty}^{\infty}\mathrm{d}t\ K\left(  t\right)  \ ,
\end{equation}

or in terms of the spectral density $P_{\phi_{\mathrm{LO}}}\left(
\omega\right)  $ of $X_{\phi_{\mathrm{LO}}}\left(  t\right)  $%

\begin{equation}
\left\langle \left(  X_{\tau}-X_{0}\right)  ^{2}\right\rangle =\frac{2\pi
}{\tau}P_{\phi_{\mathrm{LO}}}\left(  0\right)  \ . \label{Var(X)}%
\end{equation}

The responsivity $R$ of the detection scheme is defined as%

\begin{equation}
R=\left\vert \frac{\partial X_{0}}{\partial m}\right\vert \ .
\end{equation}

Using Eq. (\ref{Var(X)}) one finds that the minimum detectable change in mass
is given by%

\begin{equation}
\delta m=R^{-1}\left(  \frac{2\pi}{\tau}\right)  ^{1/2}P_{\phi_{\mathrm{LO}}%
}^{1/2}\left(  0\right)  \ .
\end{equation}

Moreover, since $\omega_{0}$ is expected to be proportional to $m^{-1/2}$ one has%

\begin{equation}
\frac{\delta m}{m}=\frac{2}{\omega_{0}}\left(  \frac{2\pi}{\tau}\right)
^{1/2}\left\vert \frac{\partial X_{0}}{\partial\omega_{0}}\right\vert
^{-1}P_{\phi_{\mathrm{LO}}}^{1/2}\left(  0\right)  \ . \label{delta M}%
\end{equation}

\section{Spectral Density}

To calculate the spectral density $P_{\phi_{\mathrm{LO}}}\left(
\omega\right)  $ of $X_{\phi_{\mathrm{LO}}}\left(  t\right)  $ it is
convenient to introduce the Fourier transform%
\begin{equation}
c(t)=\frac{1}{\sqrt{2\pi}}\int_{-\infty}^{\infty}\mathrm{d}\omega c\left(
\omega\right)  e^{-i\omega t}\ ,
\end{equation}

\begin{equation}
\Gamma(t)=\frac{1}{\sqrt{2\pi}}\int_{-\infty}^{\infty}\mathrm{d}\omega
\Gamma\left(  \omega\right)  e^{-i\omega t}\ .
\end{equation}

Assuming the bath modes are in thermal equilibrium, one finds%

\begin{equation}
\left\langle F\left(  \tau\right)  \right\rangle =\left\langle F^{\dagger
}\left(  \tau\right)  \right\rangle =0\ , \label{<F>=<F+>=0}%
\end{equation}%
\begin{equation}
\left\langle F\left(  \tau\right)  F\left(  \tau^{\prime}\right)
\right\rangle =\left\langle F^{\dagger}\left(  \tau\right)  F^{\dagger}\left(
\tau^{\prime}\right)  \right\rangle =0\ , \label{<FF>=<F+F+>}%
\end{equation}%
\begin{equation}
\left\langle F\left(  \tau\right)  F^{\dagger}\left(  \tau^{\prime}\right)
\right\rangle =\left(  \lambda_{0}+\lambda_{1}\right)  \delta\left(  \tau
-\tau^{\prime}\right)  \left\langle n_{\omega_{0}}\right\rangle \ ,
\label{<FF+>}%
\end{equation}%
\begin{equation}
\left\langle F^{\dagger}\left(  \tau\right)  F\left(  \tau^{\prime}\right)
\right\rangle =\left(  \lambda_{0}+\lambda_{1}\right)  \delta\left(  \tau
-\tau^{\prime}\right)  \left(  \left\langle n_{\omega_{0}}\right\rangle
+1\right)  \ . \label{<F+F>}%
\end{equation}

where
\begin{equation}
\left\langle n_{\omega}\right\rangle =\frac{1}{e^{\beta\hbar\omega}-1}\ ,
\end{equation}
and $\beta=1/k_{B}T$.

In Ref. \cite{Squeezing_Yurke05,Buke_023815} we have found that the following holds%

\begin{equation}
c(\omega)=\frac{\Gamma(\omega)}{(-i\omega+\lambda_{0})(-i\omega+\lambda_{1}%
)}\ . \label{a(omega)}%
\end{equation}

where%

\begin{equation}
\left\langle \Gamma(\omega)\right\rangle =\left\langle \Gamma^{\dag}%
(\omega)\right\rangle =0\ ,
\end{equation}

\begin{equation}
\left\langle \Gamma(\omega^{\prime})\Gamma(\omega)\right\rangle =\mathcal{N}%
_{1}\left(  \omega\right)  \delta\left(  \omega+\omega^{\prime}\right)  \ ,
\end{equation}

\begin{equation}
\left\langle \Gamma^{\dag}(\omega^{\prime})\Gamma^{\dag}(\omega)\right\rangle
=\mathcal{N}_{1}^{\ast}\left(  \omega\right)  \delta\left(  \omega
+\omega^{\prime}\right)  \ ,
\end{equation}

\begin{equation}
\left\langle \Gamma^{\dag}(\omega^{\prime})\Gamma(\omega)\right\rangle
+\left\langle \Gamma(\omega^{\prime})\Gamma^{\dag}(\omega)\right\rangle
=\mathcal{N}_{2}\left(  \omega\right)  \delta\left(  \omega-\omega^{\prime
}\right)  \ ,
\end{equation}

and%

\begin{equation}
\mathcal{N}_{1}\left(  \omega\right)  =2W^{\prime}W^{\ast}V\coth\frac
{\beta\hbar\omega_{0}}{2}\ ,
\end{equation}

\begin{equation}
\mathcal{N}_{2}=2W^{\prime}\left(  \left\vert W+i\omega\right\vert
^{2}+\left\vert V\right\vert ^{2}\right)  \coth\frac{\beta\hbar\omega_{0}}%
{2}\ .
\end{equation}

The frequency auto-correlation function of $X_{\phi_{\mathrm{LO}}}$ is related
to the spectral density $P_{\phi_{\mathrm{LO}}}\left(  \omega\right)  $ by%

\begin{equation}
\left\langle X_{\phi_{\mathrm{LO}}}(\omega^{\prime})X_{\phi_{\mathrm{LO}}%
}(\omega)\right\rangle =P_{\phi_{\mathrm{LO}}}\left(  \omega\right)
\delta\left(  \omega-\omega^{\prime}\right)  \ ,
\end{equation}

thus one finds%

\begin{align}
P_{\phi_{\mathrm{LO}}}\left(  \omega\right)   &  =\frac{e^{2i\phi
_{\mathrm{LO}}}\mathcal{N}_{1}\left(  \omega\right)  }{(i\omega+\lambda
_{0})(i\omega+\lambda_{1})(-i\omega+\lambda_{0})(-i\omega+\lambda_{1}%
)}\nonumber\\
&  +\frac{e^{-2i\phi_{\mathrm{LO}}}\mathcal{N}_{1}^{\ast}\left(
\omega\right)  }{(-i\omega+\lambda_{0}^{\ast})(-i\omega+\lambda_{1}^{\ast
})(i\omega+\lambda_{0}^{\ast})(i\omega+\lambda_{1}^{\ast})}\nonumber\\
&  +\frac{\mathcal{N}_{2}\left(  \omega\right)  }{(i\omega+\lambda_{0}^{\ast
})(i\omega+\lambda_{1}^{\ast})(-i\omega+\lambda_{0})(-i\omega+\lambda_{1}%
)}\ ,\nonumber\\
&
\end{align}

or in terms of the factors $W$ and $V$%

\begin{align}
P_{\phi_{\mathrm{LO}}}\left(  \omega\right)   &  =\frac{e^{2i\phi
_{\mathrm{LO}}}W^{\ast}V+e^{-2i\phi_{\mathrm{LO}}}WV^{\ast}+\left\vert
W+i\omega\right\vert ^{2}+\left\vert V\right\vert ^{2}}{(\omega-i\lambda
_{0})(\omega+i\lambda_{0})(\omega-i\lambda_{1})(\omega+i\lambda_{1}%
)}\nonumber\\
&  \times2W^{\prime}\coth\frac{\beta\hbar\omega}{2}\ .\nonumber\\
&  \label{P_phi(omega)}%
\end{align}

\subsubsection{Spectral Density at $\omega=0$}

At frequency $\omega=0$ one finds%

\begin{equation}
P_{\phi_{\mathrm{LO}}}\left(  0\right)  =\frac{1+2\zeta\cos\left(
\phi_{\mathrm{LO}}-\phi_{0}\right)  +\zeta^{2}}{\left(  1-\zeta^{2}\right)
^{2}}\frac{2W^{\prime}}{|W|^{2}}\coth\frac{\beta\hbar\omega_{0}}{2}\ ,
\label{P_phi(0)}%
\end{equation}

where the phase factor $\phi_{0}$ is defined in Eq. (\ref{exp(2i phi)}).

The largest value%

\begin{equation}
\left[  P_{\phi}\left(  0\right)  \right]  _{\max}=\frac{1}{\left(
1-\zeta\right)  ^{2}}\frac{2W^{\prime}}{|W|^{2}}\coth\frac{\beta\hbar
\omega_{0}}{2}\ ,
\end{equation}

is obtained when $\cos\left(  \phi_{\mathrm{LO}}-\phi_{0}\right)  =1$, and the
smallest value%

\begin{equation}
\left[  P_{\phi}\left(  0\right)  \right]  _{\min}=\frac{1}{\left(
1+\zeta\right)  ^{2}}\frac{2W^{\prime}}{|W|^{2}}\coth\frac{\beta\hbar
\omega_{0}}{2}\ ,
\end{equation}

when $\cos\left(  \phi_{\mathrm{LO}}-\phi_{0}\right)  =-1$.

\subsubsection{Integrated Spectral Density}

The integral over all frequencies of the spectral density is easily calculated
by employing the residue theorem%

\begin{align}
&  \frac{\int_{-\infty}^{\infty}P_{\phi_{\mathrm{LO}}}\left(  \omega\right)
\mathrm{d}\omega}{2\pi W^{\prime}\coth\frac{\beta\hbar\omega_{0}}{2}}%
=\frac{e^{2i\phi_{\mathrm{LO}}}W^{\ast}V+e^{-2i\phi_{\mathrm{LO}}}WV^{\ast
}+2\left\vert W\right\vert ^{2}}{\lambda_{0}\lambda_{1}(\lambda_{0}%
+\lambda_{1})}\ .\nonumber\\
&
\end{align}

Using Eqs. (\ref{lam_0+lam_1}) and (\ref{lam_0*lam_1}) one finds%

\begin{equation}
\frac{1}{2\pi}\int_{-\infty}^{\infty}P_{\phi_{\mathrm{LO}}}\left(
\omega\right)  \mathrm{d}\omega=\frac{1+\zeta\cos\left(  \phi_{\mathrm{LO}%
}-\phi_{0}\right)  }{1-\zeta^{2}}\coth\frac{\beta\hbar\omega_{0}}{2}\ .
\end{equation}

Thus, the integrated spectral density peaks and deeps simultaneously with
$P_{\phi_{\mathrm{LO}}}\left(  0\right)  $.

\section{Minimum Detectable Mass}

To evaluate $\delta m$ using Eq. (\ref{delta M}) the responsivity factor
$\partial X_{0}/\partial\omega_{0}$ has to be determined. Consider a small
change $\delta\omega_{0}$ in the resonance frequency. Let $c_{m}$ be the
resultant change in the steady state amplitude $C_{m}$ (here $c_{m}$ is
considered as a c-number). Using Eqs. (\ref{Theta(C_m,C_m^*)}), (\ref{W}), and
(\ref{V}) one finds%

\begin{equation}
-iC_{m}\left(  \delta\omega_{0}\right)  =Wc_{m}+Vc_{m}^{\ast}\ .
\label{-iC_m delta omega_0}%
\end{equation}

Employing a coordinate transformation to the local principal axes (see
appendix) and using Eq. (\ref{Wz+Vz* prin axes}) one finds%

\begin{equation}
\left\vert C_{m}\right\vert e^{i\phi_{C}}\left(  \delta\omega_{0}\right)
=\left[  \left(  \left\vert W\right\vert +\left\vert V\right\vert \right)
\xi+i\left(  \left\vert W\right\vert -\left\vert V\right\vert \right)
\eta\right]  \ ,
\end{equation}

where%

\begin{equation}
\phi_{C}=\phi_{m}-\phi_{a}-\pi/2\ ,
\end{equation}

and the phase factor $\phi_{m}$ is defined by Eq.
(\ref{C_m=E^(1/2)*exp(i*phi_m)}). The inverse transformation Eq. (\ref{z z*})
and Eq. (\ref{exp(2i phi)}) yield%

\begin{equation}
c_{m}=e^{-i\phi_{0}}\left\vert \frac{C_{m}}{W}\right\vert \left(  \frac
{\cos\phi_{C}}{1+\zeta}+\frac{i\sin\phi_{C}}{1-\zeta}\right)  \left(
\delta\omega_{0}\right)  \ ,
\end{equation}

or%

\begin{equation}
c_{m}=e^{-i\phi_{0}}\left\vert \frac{C_{m}}{W}\right\vert \frac{e^{i\phi_{C}%
}-\zeta e^{-i\phi_{C}}}{1-\zeta^{2}}\left(  \delta\omega_{0}\right)  \ .
\end{equation}

The change in $X_{0}$ is given by $\delta X_{0}=e^{i\phi_{\mathrm{LO}}}%
c_{m}+e^{-i\phi_{\mathrm{LO}}}c_{m}^{\ast}$, thus one has%

\begin{equation}
\frac{\partial X_{0}}{\partial\omega_{0}}=2\left\vert \frac{C_{m}}%
{W}\right\vert \operatorname{Re}\left(  e^{i\left(  \phi_{\mathrm{LO}}%
-\phi_{0}+\phi_{C}\right)  }\frac{1-\zeta e^{-2i\phi_{C}}}{1-\zeta^{2}%
}\right)  \ . \label{dX_0/d omega_0}%
\end{equation}

Finally, using Eqs. (\ref{delta M}), (\ref{P_phi(0)}), and
(\ref{dX_0/d omega_0}), and assuming the case of high temperature%

\begin{equation}
\beta\hbar\omega_{0}\ll1\ ,
\end{equation}

one finds%

\begin{equation}
\frac{\delta m}{m}=2\left(  \frac{2\pi}{Q_{\mathrm{eff}}\omega_{0}\tau}%
\frac{k_{B}T}{U_{0}}\right)  ^{1/2}g\left(  \phi_{\mathrm{LO}}-\phi
_{0}\right)  \ , \label{delta_m/m nonlinear}%
\end{equation}

where $Q_{\mathrm{eff}}=\omega_{0}/W^{\prime}$ is the effective quality
factor, the function $g$ is given by%

\begin{equation}
g\left(  \phi\right)  =\frac{\left[  1+2\zeta\cos\phi+\zeta^{2}\right]
^{1/2}}{\left\vert \cos\left(  \phi+\phi_{C}\right)  -\zeta\cos\left(
\phi-\phi_{C}\right)  \right\vert }\ ,
\end{equation}

and%

\begin{equation}
U_{0}=\hbar\omega_{0}\left\vert C_{m}\right\vert ^{2}\ .
\end{equation}

\begin{figure}
[ptb]
\begin{center}
\includegraphics[
height=4.7124in,
width=3.243in
]%
{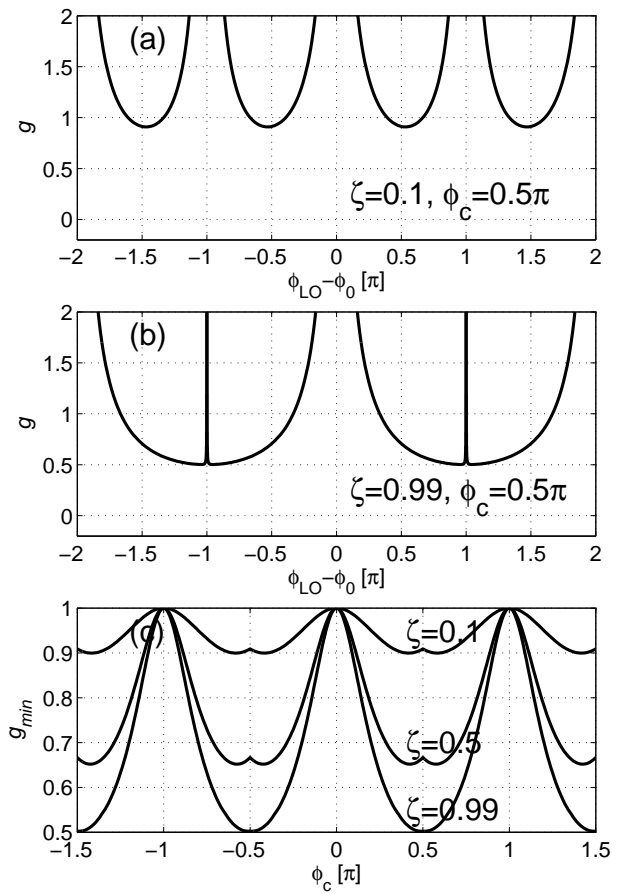}%
\caption{The function $g$. Panel (a) shows $g\left(  \phi_{\mathrm{LO}}%
-\phi_{0}\right)  $ for the case $\zeta=0.1$ and $\phi_{C}=0.5\pi$, and panel
(b) for the case $\zeta=0.99$ and $\phi_{C}=0.5\pi$. Panel (c) shows the
minimum value of the function $g\left(  \phi_{\mathrm{LO}}-\phi_{0}\right)  $
vs. $\phi_{C}$ for different values of $\zeta$.}%
\label{g_function}%
\end{center}
\end{figure}

In view of a comparison between Eq. (\ref{delta m linear}) and Eq.
(\ref{delta_m/m nonlinear}) we refer to the case where $g<1$ as the case where
the lower bound imposed upon the minimum detectable mass of a linear resonator
is exceeded. The function $g\left(  \phi_{\mathrm{LO}}-\phi_{0}\right)  $ is
plotted in Fig. (\ref{g_function}) (a) for the case $\zeta=0.1$ and $\phi
_{C}=0.5\pi$, and in Fig. (\ref{g_function}) (b) for the case $\zeta=0.99$ and
$\phi_{C}=0.5\pi$. For both cases values of $g$ below unity are obtained in
some range of $\phi_{\mathrm{LO}}$. Figure (\ref{g_function}) (c) shows the
minimum value of the function $g\left(  \phi_{\mathrm{LO}}-\phi_{0}\right)  $
vs. $\phi_{C}$ for 3 different values of $\zeta$. In general $0.5\leq g_{\min
}\leq1$ for all values of $\phi_{C}$ and $\zeta$, whereas, the lowest value
$g_{\min}=0.5$ is obtained in the limit $\zeta\rightarrow1$. This limit
corresponds to the case of operating close to a jump point, namely close to
the edge of the bistability region.%

\section{Conclusions}

In the present paper we analyze the performances of a nanomechanical mass
detector. Both Kerr nonlinearity and nonlinear damping are taken into account
to lowest order. The lower bound imposed upon the minimum detectable mass due
to thermomechanical noise is generalized for the present case. The lowest
detectable mass is obtained when the resonator is driven close to a jump point
near the edge of the bistability region. However, in the same region
slowing-down occurs in the response of the detector to a change in the mass
(see Eq. (\ref{t_RD nonlinear})), limiting thus the detection speed. In
general, for a given application the operating point can be chosen to
optimally balance between the different requirements on the sensitivity and
response time.

\appendix

\section{Principal Axes}

Consider an expansion of the function $\Theta$ near a complex number $Z$%

\begin{equation}
\Theta\left(  Z+z,Z^{\ast}+z^{\ast}\right)  =\Theta_{0}+Wz+Vz^{\ast}+O\left(
\left\vert z\right\vert ^{2}\right)  \ ,
\end{equation}

where $\Theta_{0}=\Theta_{0}\left(  Z,Z^{\ast}\right)  $, and $W$ and $V$ are
given by Eqs. (\ref{W}) and (\ref{V}) respectively.

The transformation%

\begin{equation}
\left(
\begin{array}
[c]{c}%
\xi\\
\eta
\end{array}
\right)  =\frac{1}{2}\left(
\begin{array}
[c]{cc}%
e^{i\phi} & e^{-i\phi}\\
-ie^{i\phi} & ie^{-i\phi}%
\end{array}
\right)  \left(
\begin{array}
[c]{c}%
z\\
z^{\ast}%
\end{array}
\right)  \ , \label{(xi,eta)}%
\end{equation}

represents axes rotation with angle $\phi$ ($\phi$ is real). The inverse
transformation is given by%

\begin{equation}
\left(
\begin{array}
[c]{c}%
z\\
z^{\ast}%
\end{array}
\right)  =\left(
\begin{array}
[c]{cc}%
e^{-i\phi} & ie^{-i\phi}\\
e^{i\phi} & -ie^{i\phi}%
\end{array}
\right)  \left(
\begin{array}
[c]{c}%
\xi\\
\eta
\end{array}
\right)  \ . \label{z z*}%
\end{equation}

Using this notation one finds%

\begin{equation}
Wz+Vz^{\ast}=R_{\xi}\xi+R_{\eta}\eta\ ,
\end{equation}

where%

\begin{align}
R_{\xi}  &  =We^{-i\phi}+Ve^{i\phi}\ ,\\
R_{\eta}  &  =i\left(  We^{-i\phi}-Ve^{i\phi}\right)  \ .
\end{align}

Principle axes are obtained by choosing $\phi=\phi_{0}$ where%
\begin{equation}
e^{2i\phi_{0}}=\frac{WV^{\ast}}{\left\vert WV\right\vert }\ .
\label{exp(2i phi)}%
\end{equation}

Thus, using the notation%

\begin{equation}
\left(  \frac{WV}{\left\vert WV\right\vert }\right)  ^{1/2}=e^{i\phi_{a}}\ ,
\label{exp(i*phi_a)}%
\end{equation}

one finds that in the reference frame of the principle axes the following hold%

\begin{align}
R_{\xi}  &  =e^{i\phi_{a}}\left(  \left\vert W\right\vert +\left\vert
V\right\vert \right)  \ ,\label{R_xi}\\
R_{\eta}  &  =ie^{i\phi_{a}}\left(  \left\vert W\right\vert -\left\vert
V\right\vert \right)  \ , \label{R_eta}%
\end{align}

and%

\begin{equation}
Wz+Vz^{\ast}=e^{i\phi_{a}}\left[  \left(  \left\vert W\right\vert +\left\vert
V\right\vert \right)  \xi+i\left(  \left\vert W\right\vert -\left\vert
V\right\vert \right)  \eta\right]  \ . \label{Wz+Vz* prin axes}%
\end{equation}


\section*{Acknowledgment}

This work is supported by the Israeli ministry of science, Intel Corp.,
Israel-US binational science foundation, and by Henry Gutwirth foundation.

\newpage
\bibliographystyle{apsrev}
\bibliography{acompat,Eyal_Bib}

\end{document}